\documentclass[preprint,showpacs,preprintnumbers,amsmath,amssymb]{revtex4}
\usepackage{graphicx}
\usepackage{dcolumn}
\usepackage{bm}
\begin{document}
\title{Mass Fractal Dimension of the Ribosome and Implication of
  its Dynamic Characteristics}
\author{Chang-Yong Lee}
\email{clee@kongju.ac.kr}
\affiliation{The Department of Industrial Information, Kongju National
  University, Chungnam, 340-702 South Korea}
\date{\today}

\begin{abstract}
Self-similar properties of the ribosome in terms of the mass
fractal dimension are investigated. We find that both the 30S subunit
and the 16S rRNA have fractal dimensions of 2.58 and 2.82,
respectively; while the 50S subunit as well as the 23S rRNA has the
mass fractal dimension close to 3, implying a compact three dimensional 
macromolecule. This finding supports the dynamic and active
role of the 30S subunit in the protein synthesis, in contrast to the
pass role of the 50S subunit.
\end{abstract}


\maketitle
The structure of biomolecules is important because not only the
structure dictates its biological function, but it is the target of
antibiotics. In this sense, finding characteristics of the three-dimensional
shape of biomolecules is important for a better understanding of their
biological functions and associated applications in medicine. In the
case of the ribosome~\cite{ribosome}, as a large protein-RNA complex,
in contrast to most cellular machines, it has been known that the
ribosomal function heavily rely on the ribosomal RNA (rNRA), as a
ribozyme, rather than protein components~\cite{function}. In
particular, the protein synthesis is closely related to the dynamic
structure of the ribosome, which is too complicated for direct studies
by physical methods. However, careful study on the static structure from a
quantitative perspective may reveal an important aspect of its dynamic
properties.  

The quantitative investigation on the structure of the ribosome is
relatively less studied than that of protein, mainly due to the
difficulty of the highly resolved structure conformation. In contrast
to the ribosome, the geometric and self-similar properties of the
proteins have been studied extensively. It was reported that the
relation between the average radius and the mass of the protein chains
can be described by a fractal dimension~\cite{moret}. The mass fractal
dimension of the protein was shown to lie near 3, suggesting a compact
three-dimensional object~\cite{elber}. However, more recent and
extensive studies, including a statistical analysis in estimating the
dimension, argue smaller values that are consistent with the
result of the vibrational analysis of proteins~\cite{enright}. These
results suggest that the fractal dimension of less than 3
may be intrinsic and universal characteristics of the protein chain.  

Since the ribosome, as a tightly packed macromolecule, was considered
too large for a high-resolution structural analysis, quantitative
studies toward an understanding of the structure proved difficult
until recent progress in the high resolution crystallography has been
made. In fact, the ribosome and its subunits are the largest
asymmetric molecules that have been resolved at the atomic level so far
by the crystallography. The 2.4 $\mathring{A}$ high-resolution of the
50S subunit from the {\it Haloarcula marismortuii}~\cite{ban} and the
3.05 $\mathring{A}$ structure of the 30S subunit from the {\it
  Thermus thermophilus}~\cite{wimberly} provided the first detailed
views of the structure of both ribosomal subunits; the intact 70S
ribosome from a {\it Escherichia coli} of 3.5 $\mathring{A}$
resolution~\cite{schuwirth} revealed the features of the inter-subunit
bridges. In addition to these, there are other X-ray crystal
structures available for the ribosome as well as its
subunits~\cite{harms,schluenzen,cate,yusupov}. With these considerable
advances in the ribosome structures at the atomic level, we are now
able to investigate quantitative characteristics of the ribosome from
the statistical physics perspective.    

In bacteria, the ribosome is a particle of size about 250
$\mathring{A}$ in diameter and consists mainly of two subunits: 30S and 50S
subunits, together forming the 70S. The unit ``S'' stands for Svedberg,
which is a measure of the sedimentation rate. The 30S subunit plays a
crucial role in decoding mRNA by monitoring base pairing between codon
and anticodon; whereas the 50S subunit catalyzes peptide bond
formation between the incoming amino acid and the nascent peptide
chain~\cite{ramakrishnan}. The 30S subunit, in turn, contains the 16S
rRNA molecule in addition to about 20 different proteins, and the 50S subunit 
consists of the 5S and the 23S rRNAs besides about 30 different
proteins. The 16S and 23S rRNAs are composed of approximately 1500 and
3000 nucleotides respectively, each of which is composed of one of four 
different bases (denoted as A, C, G, and U), and sugar-phosphate backbones. 

With this structural information of the ribosome at the atomic level,
we investigate the self-similar property of the ribosome structure and
its biological implication. We especially focus on the scale
invariance, by estimating the mass fractal dimension, for structures of
{\it Thermus thermophilus} 30S subunit including the 16S rRNA, and
{\it Haloarcula marismortui} 50S subunit including the 23S
rRNA~\cite{data}. It has been known that the protein synthesis occurs
in the context of the intact ribosome and the moving part of the
ribosome enables the dynamic process of the translation. In this
sense, the ribosome function is closely related to its spatial
conformation in the physiological medium. Thus, the mass fractal
dimension analysis may help to reveal any characteristics of the ribosome,
especially the dynamics of the ribosome. 

The mass fractal dimension, which can be used as a measure of the compactness,
is defined as the number of monomers (atoms in our case), $N$,
enclosed in a sphere of radius $R$. When a molecule has a fractal
structure, it is expected that  
\begin{equation}
N \propto R^{D_{M}}~,
\end{equation}
where $D_{M}$ is the mass fractal dimension. It can be estimated by
plotting the number of all atoms contained inside concentric spheres
of varying radius $R$ on a log-log scale. 

To test the sensitivity of the result to the choice of the origin, we
set the origin at the geometric center of each molecule, 
and vary the origin by $\pm 2~\mathring{A}$ in the XYZ directions
(thus 27 different origins) with respect to the coordinate system
adopted from the PDB data. We ignore hydrogen atoms since the X-ray crystal
structures do not contain the geometric information of hydrogen atoms,
which cannot be seen but in very high resolution. Incidentally, this
is also true for the protein case. Thus, most of descriptions of the 
ribosome focus on the position of the heavy atoms, such as C, N, O,
and P. Note also that the PDB for the 30S
subunit does not contain water molecules; where as that for the 50S
does. For a fair comparison, we exclude water molecules from the mass
fractal calculation. 
Due to the finite-size effect, there are both upper and lower size
limits beyond which a macromolecule is no longer fractal.

\begin{table}
\begin{ruledtabular}
\caption{The number of atoms, the average, and its standard deviation
  of the mass fractal dimension for each  molecule. For the 16S rRNA
  and the 30S subunit, the scaling property emerges for all 27 measurements;
  while for the 23S rRNA and the 50S subunit, respectively 17 and 19
  out of 27 measurements show scaling properties. The parentheses are
  the standard deviation of the corresponding number of estimations.}
\setlength{\extrarowheight}{4pt}
\begin{tabular}{c c c}
molecule  &  number of atoms  & $D_{M}$     \\ \hline 
16S rRNA    & 32514           & 2.58 (0.06) \\
30S subunit & 51742           & 2.82 (0.07) \\
23S rRNA    & 59017           & 3.11 (0.07) \\
50S subunit & 62673           & 3.07 (0.08) \\
\end{tabular}
\end{ruledtabular}
\end{table}

We estimate the mass fractal dimension for the 16S rRNA, the 30S subunit,
the 23S rRNA, and the 50S subunit with 27 different origins. The
result is summarized in Table~1, and Fig.~1 and 2 present a typical
log-log plot of the enclosed ``mass'' $N$ as a function of radius $R$. 
Note that the geometric origin and the center of mass for all molecules
are almost identical, and evaluating the corresponding physical 
mass rather than the number of atoms does not affect the result.
For the 23S rRNA and the 30S subunit, we are able to estimate the
fractal dimensions for all 27 cases; while for the 23S rRNA and the
50S subunit, respectively 17 and 19 cases out of 27 shows the scaling
behavior. 

The mass fractal dimension exhibits the molecule's space-filling
ability: the larger is $D_{M}$, the more atoms are in the sphere. When
fractal dimension is less than 3, the structure has ``empty'' or
``void'' space. From the above result, we see that the mass fractal
dimensions for both the 23S rRNA and the 50S subunit are close to 3,
implying that these are compact three-dimensional collapsed
objects. On the other hand, the average $D_{M}$ for the 16S rRNA and
the 30S subunit are found to be 2.58 and 2.82, respectively, smaller
than that of a completely compact three-dimensional collapsed polymer.
This indicates that the mass inside a radius $R$ does not increase with
the Euclidean dimension as an exponent but with some lesser
power. Thus, we find that both the 23S rRNA and 50S subunit are more 
compact than either the 16S rRNA or the 30S subunit. 

The fact that the 16S rRNA and the 30S subunit have the mass fractal
dimension less than 3 leaves a room for the 16S rRNA  to
make any movement during the protein synthesis, and can be related to
the rigid body motion of domains within the subunit.
It has been found from many studies that it is the 30S subunit that
makes a motion in translocation. It was reported that the
30S subunit makes a ratchet-like rotations relative to the large
50S subunit~\cite{schuwirth,frank}, particularly, the
rotational rigid body motions of the ``head'' domain~\cite{domain} 
within the 30S subunit~\cite{schuwirth}. This reveals a high degree of
flexibility between the head and the rest of the 30S subunit. 
Furthermore, not only domains for the entrance channel such as the
``shoulder'' and the head domains in the 30S subunit are dynamic
during decoding, but the exit channel formed between the 
``platform'' and the head domains are known to be
variable~\cite{schuwirth,frank,spirin,serdyuk}. Thus, it is the 30S
subunit that is either dynamic or variable, playing an active role,
which is possible due to the fractal characteristics of the 30S subunit. 

The 50S subunit, on the other hand, cannot make movements due
to a highly compact structure except peripheral regions. As an
example, the L1 stalk, located in a peripheral region of the 50S
subunit, makes a bifurcation~\cite{frank} and moves toward the
inter-subunit space playing a pivotal role in the transcription
process~\cite{valle}. 

In this paper, we investigated a symmetry embedded in the ribosome structure
under a change of the length scale via the mass fractal dimension. We
found the 30S and the 50S subunit (also the 16S and the 23S rRNAs)
differs in their fractal dimensions: the 30S subunit and the 16S rRNA
have fractal dimensions, while the 50S subunit and the 23S rRNA can be
regarded as three-dimensional compact molecules. The fractality of
both the 16S rRNA and the 30S subunit supports the dynamic nature of
the ribosome in the protein synthesis. 

Although the power of the self-similarity approach to the ribosome
structure is in its simplicity and generality, it also true that their
detail dynamic properties and realization are not obvious because
detailed properties of the ribosome which determine their function
are averaged out. Nevertheless, the fractal property of the 30S
subunit (also the 16S rRNA) provides a partial, if not whole, evidence of
its movement during the transcription. 

This work was supported by the Korea Research Foundation Grant funded
by the Korean Government (MOEHRD) (KRF-2005-041-H00052). 

\newpage
\begin{figure}
\centerline{
\includegraphics{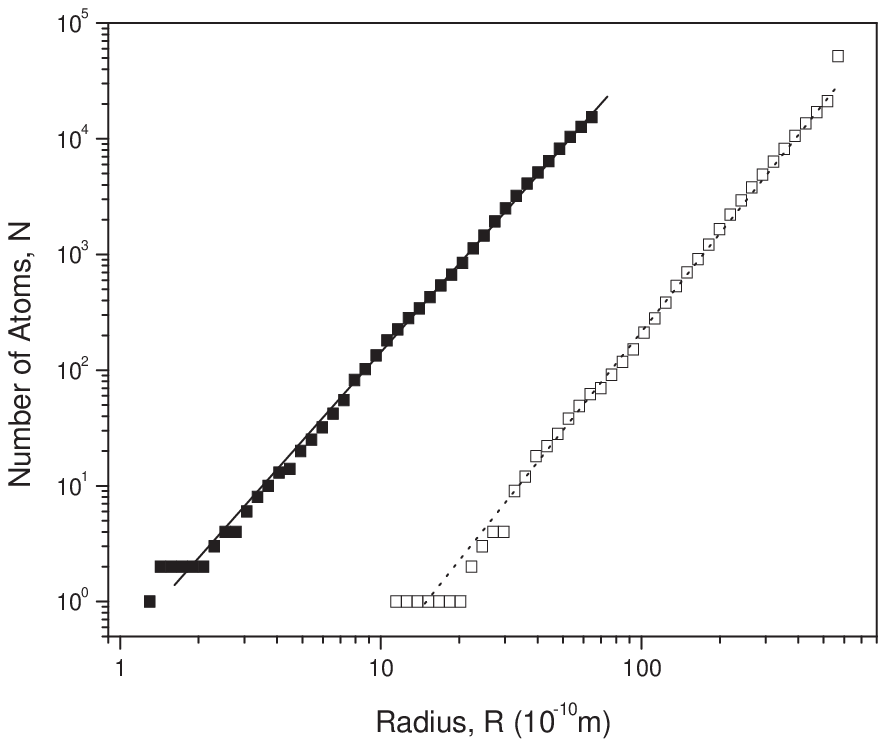}
}
\caption{A typical log-log plot of the number of atoms versus the
  radius for the 16S rRNA ($\blacksquare$) and the 30S subunit
  ($\Box$). The least-square fits on the slope of the 16S rRNA (solid
  line) and the 30S subunit (dotted line) yield 2.62 and 2.85,
  respectively. The plot for the 30S subunit is shifted to the 
  right by the factor of 10 for the display purpose. }
\end{figure}
\begin{figure}
\centerline{
\includegraphics{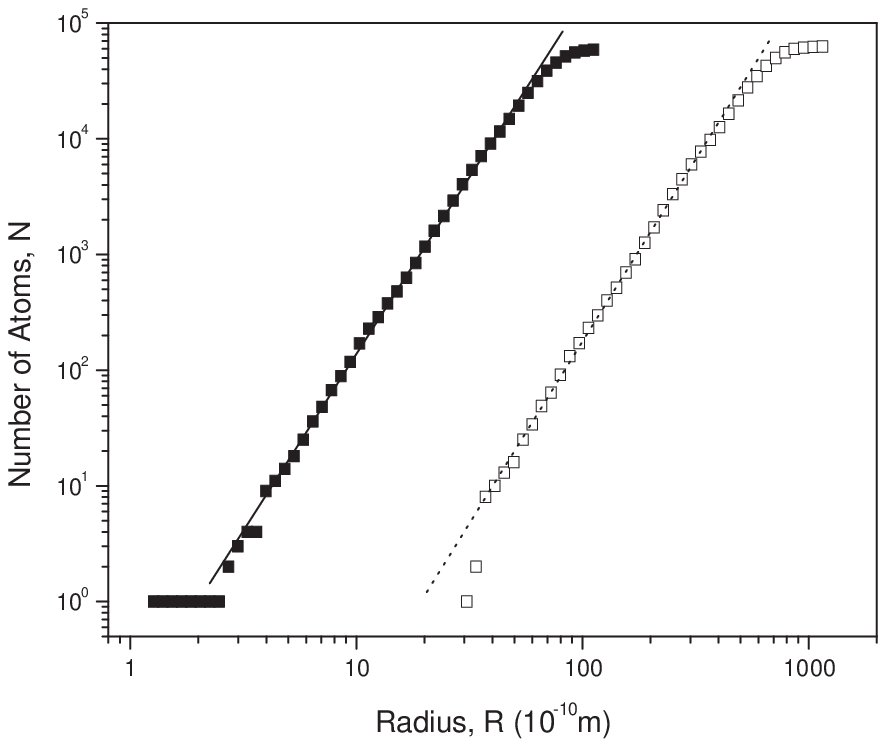}
}
\caption{A typical log-log plot of the number of atoms versus the
  radius for the 23S rRNA ($\blacksquare$) and the 50S subunit
  ($\Box$). The least-square fits on the slope of the 23S rRNA (solid
  line) and the 50S subunit (dotted line) yield 3.01 and 3.06,
  respectively. The plot for the 50S subunit is shifted to the 
  right by the factor of 10 for the display purpose.}
\end{figure}

\begin{thebibliography}{99}
\bibitem{domain} The structure of the 16S rRNA is commonly organized
  into four domains of a few hundred nucleotides each: the head, the body, and
  the platform, and $3^{\prime}$ minor domains.
\end{thebibliography}
\end{document}